\documentclass[12pt,a4paper]{article}

\newcommand{\g}{\gamma}
\newcommand{\G}{\Gamma}
\newcommand{\F}{\mathcal{F}}
\newcommand{\p}{\partial}
\newcommand{\ta}{\theta}
\newcommand{\la}{\lambda}
\begin{document}

\begin{flushright}
hep-th/9809088 
\end{flushright}
\vspace{1.8cm}

\begin{center}
 \textbf{\Large Covariant Gauge Fixing for Super IIA D-branes} 
\end{center} 
\vspace{1.6cm}
\begin{center} 
  Shijong Ryang 
\end{center}

\begin{center}
\textit{Department of Physics \\ Kyoto Prefectural  University of Medicine 
\\ Taishogun, Kyoto 603-8334 Japan}
 \par
\texttt{ryang@koto.kpu-m.ac.jp}
\end{center}
\vspace{2.8cm}
\begin{abstract}
For the $\kappa$-symmetric super IIA D-brane action by the canonical 
approach we construct an equivalent effective action which is characterized
by an auxiliary scalar field. By analyzing the canonical equations of 
motion for the $\kappa$-symmetry-gauge-fixed action we find a suitable
conformal-like covariant gauge fixing of reparametrization symmetry to 
obtain a simplified effective action where the non-linear square root
structure is removed. We discuss how the two effective actions are
connected.
\end{abstract}
\vspace{3cm}
\begin{flushleft}
 September, 1998
\end{flushleft}
\newpage
The emergence of D-branes in the superstring theory has led to remarkable
developments in our understanding of the non-perturbative aspects of the
theory. Since the R-R p-brane solutions of type II supergravity theories
are interpreted as the D-p-branes which are described by the worldvolumes
on which the open superstring ends \cite{PCJ}, their bosonic worldvolume
action can be constructed by the usual superstring methods as the sum
of a Born-Infeld-Nambu-Goto (BING) action \cite{FT,RL} and a Wess-Zumino
(WZ) action \cite{ML,MD,CS,AT,GHT}. The global supersymmetry and local 
kappa-symmetry invariant actions for the D-p-branes have been constructed 
in flat superspace \cite{APS,APPS} and the general type IIA or IIB 
supergravity background \cite{CGNW,BT1}. 

From the bosonic D-p-brane non-linear BING action an equivalent effective
action in a sort of Polyakov type 
which is quadratic in the derivative of 
ten-dimensional space-time coordinates and linear in the Born-Infeld U(1) 
gauge field strength, has been produced by means of the Lagrangian 
description \cite{ZH} where the extension to super D-p-brane is also 
presented, and the canonical phase space description \cite{LU} where a null
bosonic D-p-brane action is also derived. Based on this quadratic effective
action for the super D-p-brane, which includes, however, an auxiliary 
'metric' with both symmetric and anti-symmetric parts, a light-cone 
Hamiltonian which is quadratic in canonical momenta has been constructed 
\cite{JL}, while a light-cone Hamiltonian for the bosonic D-2-brane has
been obtained directly from the beginning D-2-brane non-linear BING 
action \cite{MMM}. Associated with the null D-p-brane action a new 
version of the super D-p-brane action without the 
WZ term, which is $\kappa$-symmetric
even though there is no WZ term, has been constructed \cite{BT2}, where the
tension is elevated to a dynamical variable and the corresponding extended 
Hamiltonian is presented.

Toward the covariant quantization of the super D-p-brane a covariant gauge 
for fixing the $\kappa$-symmetry has been found to make the action so simple
that the WZ term disappears and a static gauge for fixing the bosonic 
reparametrization symmetry has been chosen, however, to leave a complicated
non-linear BING action \cite{APS}. From the $\kappa$-symmetry-gauge-fixed 
action which is considered as a supersymmetric generalization of the theory
in Ref. \cite{LU}, the structure of the extended Hamiltonian has been 
analyzed without using the static gauge and the BPS states of the theory
have been described Lorentz covariantly \cite{RK}. For the 
$\kappa$-symmetric type II D-p-brane action the 
WZ term has been explicitly constructed and used to
develop the canonical formulation \cite{KH}, where the global supersymmetry
algebra is computed to give the expressions of the central charges.

It is important for making the study of super D-brane dynamics tractable to
seek for an appropriate covariant gauge that linearizes the complicated 
square root structure of the BING action, instead of the static gauge. 
Starting from the $\kappa$-symmetric IIA D-p-brane action we will 
give a Dirac Hamiltonian in 
the canonical formulation. From it an effective action parametrized with an
auxiliary scalar field will be constructed. On the other hand analyzing the
canonical equations of motion for the $\kappa$-symmetry-gauge-fixed action 
we will find a suitable covariant gauge for fixing the remaining 
reparametrization symmetry, which removes the non-linearity of the 
BING action. Then we will obtain a simple quadratic
effective action. It will be observed that there is a connection between
the two effective actions.

Let us consider the super IIA D-p-brane with tension $T$ and even p which is
described by the $\kappa$-symmetric action in flat superspace
\begin{equation}
S = S_{DBI} + S_{WZ} = T \left( -\int d^{p+1}\sigma \sqrt{-\det(\g_{ij} + 
{\F}_{ij} )} + \int \Omega_{p+1} \right).
\label{act}\end{equation}
The BING action $S_{DBI}$ is manifestly global supersymmetric since it is 
expressed in terms of the manifestly supersymmetric induced worldvolume 
metric 
\begin{eqnarray}
\g_{ij} = \eta_{\mu\nu}\Pi^{\mu}_i\Pi^{\nu}_j,&  \Pi^{\mu}_i = \p_iX^{\mu}
- \bar{\ta}\G^{\mu}\p_i\ta,
\end{eqnarray}
where $X^{\mu}(\mu = 0,\cdots,9)$ are vector coordinates in the flat 
ten-dimensional target space and $\ta$ is the type IIA Majorana spinor, and
the supersymmetric Born-Infeld field strength of the U(1) gauge field 
$A_i(i = 0,\cdots,p)$
\begin{equation}
{\F}_{ij} \equiv {F}_{ij} - b_{ij} = [ {\p}_iA_j - \bar{\ta}{\G}_{11}\G_{\mu}
\p_{i}\ta({\p}_j X^{\mu} - \frac{1}{2} \bar{\ta}\G^{\mu}\p_j\ta)] - 
(i \leftrightarrow j),
\end{equation}
where there are no background fields. The 32-component IIA spinor $\ta$ is 
decomposed into two Majorana-Weyl spinors as $\ta = \ta_1 + \ta_2$
with $\ta_1 = (1 + \G_{11})\ta/2, \ta_2 = (1 - \G_{11})\ta/2$. The WZ action
which is specified by $I_{p+2} = d\Omega_{p+1}$ is also invariant under the
global supersymmetry bcause $I_{p+2}$ is constructed from the supersymmetry
invariants and the variation of the (p+1)-form $\Omega_{p+1}$ is exact 
\cite{APS}.  The corresponding full Lagrangian $L = L_{DBI} + L_{WZ}$ is 
considered to depend on $\pi^{\mu}_{0}, \dot{\ta}_1, \dot{\ta}_2$ and
${\F}_{0a}$. For the fields of the (p+1)-dimensional worldvolume theory such
as the ten-dimesional space-time coordinates $X^{\mu}$, 
the BI U(1) gauge field 
$A_{i}$ and a pair of Majorana-Weyl spinors $\ta_1$ and $\ta_2$, the 
canonical conjugate momenta are given by
\begin{eqnarray}
P_{\mu} &=& \frac{\p L}{\p\dot{X^{\mu}}} = \tilde{P_{\mu}} + 
\frac{\p L_{WZ}}{\p\pi^{\mu}_0} + \pi^a\bar{\ta}\G_{11}\G_{\mu}\p_a\ta, 
\nonumber \\
\pi^a &=& \frac{\p L}{\p\dot{A}_a} =  \tilde{\pi}^a + \frac{\p L_{WZ}}
{\p{\F}_{0a}}, \; (a=1,\cdots,p), \hspace{1cm}\pi^0 = 
\frac{\p L}{\p\dot{A}_0} = 0
\end{eqnarray}
and
\begin{equation}
\bar{P}_{\ta_A} = \left( \tilde{P}_{\mu} + \pi^a\frac{\p{\F}_{0a}}
{\p\pi^{\mu}_0} \right)\frac{\p \pi^{\mu}_0}{\p\dot{\ta}_A}  
+ \pi^a\frac{\p{\F}_{0a}}{\p\dot{\ta}_A} + 
\frac{\p L_{WZ}}{\p\dot{\ta}_A}, \; ( A = 1,2).
\label{pth}\end{equation}
Here $\tilde{P}_{\mu}$ and $\tilde{\pi}^a$ are expressed as
\begin{eqnarray}
\tilde{P}_{\mu} &=& - \frac{T}{2} \sqrt{-\det G} G^{(0,i)}\pi_{i\mu}, 
\nonumber \\ 
\tilde{\pi}^a &=& - \frac{T}{2} \sqrt{-\det G} G^{[a,0]}
\end{eqnarray}
with $G_{ij} = \g_{ij} + {\F}_{ij}$, and satisfy the 
bosonic p+1 primary constraints
\begin{eqnarray}
T_0 &=& \frac{1}{2}( \tilde{P}^2 + \tilde{\pi}^a\hat{\g}_{ab}\tilde{\pi}^b +
T^2\det\hat{G}_{ab} ) = 0,    \nonumber \\
T_a &=& \tilde{P}_{\mu}\pi^{\mu}_a + \tilde{\pi}^b{\F}_{ab} = 0,
\end{eqnarray}
where $\hat{G}_{ab}$ is the matrix with only worldspace components.
Since $\dot{\ta}_A$ also cannot be expressed in terms of the canonical 
variables the expressions (\ref{pth}) yield the fermionic 32 constraints
\begin{eqnarray}
\bar{\Phi}_1 &=& \bar{P}_{{\ta}_1} + \bar{\ta}_1[\tilde{P}_{\mu} - \pi^a(
\pi_{a\mu} + \bar{\ta}_1\G_{\mu}\p_a\ta_1)]\G^{\mu} - \frac{\p L_{WZ}}
{\p\dot{\ta}_1} = 0, \nonumber \\
\bar{\Phi}_2 &=& \bar{P}_{{\ta}_2} + \bar{\ta}_2[\tilde{P}_{\mu} + \pi^a(
\pi_{a\mu} + \bar{\ta}_2\G_{\mu}\p_a\ta_2)]\G^{\mu} - \frac{\p L_{WZ}}
{\p\dot{\ta}_2} = 0,
\label{phi}\end{eqnarray}
where we have used the right derivatives for fermionic fields as
\begin{equation}
\frac{\p {\F}_{0a}}{\p \dot{\ta}_A} =
 (-1)^{A+1}\bar{\ta}_A{\G}_{\mu}(
{\pi}^{\mu}_a + \frac{1}{2}\bar{\ta}{\G}^{\mu}\p_a \ta) 
 + \frac{1}{2}\bar{\ta}_A {\G}^{\mu}(\bar{\ta}{\G}_{11}{\G}_{\mu}{\p}_a\ta).
\end{equation}
These 32 constraints which consist of 16 first class ones and
16 second ones include additional terms $\bar{\ta}_A\G_{\mu}\p_a\ta_A \; 
(A=1,2)$ compared to those presented in Ref.\cite{RK}, where in the first 
approximation the space-dependence on $\sigma^a$ of spinors $\ta_A$ was 
neglected. It follows that the canonical Hamiltonian vanishes identically up
to a term 
\begin{equation}
H^c = P_{\mu}\dot{X}^{\mu} + \pi^i\dot{A}_i + \sum_{A}\bar{P}_{\ta_{A}}
\dot{\ta}_A - L_{DBI} - L_{WZ} = \pi^a\p_a A_0,
\end{equation}
where $L_{WZ}$ is linear in velocities so as to satisfy
\begin{equation}
\frac{\p L_{WZ}}{\p\dot{X}^{\mu}}\dot{X}^{\mu} + \frac{\p L_{WZ}}
{\p {\F}_{0a}}{\F}_{0a} + \sum_{A}\frac{\p L_{WZ}}{\p\dot{\ta}_A}\dot{\ta}_A
= L_{WZ}.
\label{wz}\end{equation}
In this sense the super D-p-brane is a completely degenerate system.
Therefore using the Lagrange multipliers $\xi,\rho,\rho^a,\psi_1$ and 
$\psi_2$ for the constraints we construct a Dirac Hamiltonian
\begin{equation}
H = \pi^a{\p}_aA_0 + \xi \pi^0 + \rho T_0 + \rho^aT_a - \bar{\Phi}_A\psi_A,
\end{equation}
where the first class constraints such as $T_a, T_0$ and $\pi^0$ generate 
worldspace diffeomorphisms, time translations and BI gauge transformations.

Now we use the Hamiltonian to present the phase space form of the super
D-p-brane action
\begin{equation}
S_{ps} = \int d^{p+1}\sigma ( P_{\mu}\dot{X}^{\mu} + \pi^a F_{0a} + 
\bar{P}_{\ta_{A}}\dot{\ta}_{A} - \xi \pi^0 - \rho T_0 - \rho^a T_a + 
\bar{\Phi}_A\psi_A).
\end{equation}
This phase space action will be shown to be classically equivalent to the
starting super D-p-brane non-linear action (\ref{act}). The variations with
respect to the momenta $\bar{P}_{\ta_A}, P_{\mu}$ yield
\begin{eqnarray}
\psi_A &=& - \dot{\ta}_A, \nonumber \\
\tilde{P}^{\mu} &=& \frac{\dot{X}^{\mu} - \rho^a\pi^{\mu}_a + \bar{\ta}_A
\G_{\mu}\psi_A}{2\rho} = \frac{\pi^{\mu}_0 - \rho^a\pi^{\mu}_a}{2\rho}.
\label{mom}\end{eqnarray}
Varying the action with respect to the momentum $\pi^a$ and using the two
equations in (\ref{mom}) we can derive an equation expressed in terms of 
${\F}_{ij}$ 
\begin{equation}
\tilde{\pi}^a = \frac{\hat{\g}^{ab}( {\F}_{0b} - \rho^c{\F}_{cb} )}{2\rho},
\label{pai}\end{equation}
where $\hat{\g}^{ab}$ is the inverse of the matrix with worldspace 
components $\hat{\g}_{ab}$. There is a similarity between the expression of
$\tilde{P}^{\mu}$ in (\ref{mom}) and that of $\tilde{\pi}^a$. In order to
obtain the compact expression as $\tilde{\pi}^a$ in (\ref{pai}) the 
additional terms in (\ref{phi}) are necessary ingredients. Substitutions of
(\ref{mom}) and (\ref{pai}) back into $S_{ps}$ yield a configuration space
action
\begin{eqnarray}
S_{cs} &=& \int d^{p+1} \sigma [ \frac{1}{4\rho} ( \g_{00} - 2\rho^a
\g_{0a} + \rho^a\rho^b\g_{ab} + \hat{\g}^{ab}({\F}_{0a} - \rho^c{\F}_{ca})
({\F}_{0b} - \rho^d {\F}_{db}) ) \nonumber \\
&-& \rho T^2 \det {\hat{G}}_{ab} 
+ \frac{\p L_{WZ}}{\p \dot{X}^{\mu}}
\dot{X}^{\mu} + \frac{\p L_{WZ}}{\p {\F}_{0a}}{\F}_{0a} + \frac{\p L_{WZ}}
{\p \dot{\ta}_A}\dot{\ta}_A ],
\label{scs}  \end{eqnarray}
where the WZ term is properly recovered through (\ref{wz}). The bosonic 
version of (\ref{scs}) was presented in Ref.\cite{LU}, where there is no WZ
term. The stationary equation concerning the Lagrange multiplier field 
$\rho^a$ is given by
\begin{equation}
\rho^a = \hat{K}^{ab} K_{b0},
\label{raw}\end{equation}
where $K_{ij}$ is defined as $K_{ij} = \g_{ij} - {\F}_{ia}\hat{\g}^{ab}
{\F}_{bj}$ and $\hat{K}^{ab}$ is the inverse of the matrix with spatial 
components $\hat{K}_{ab}$. By the Gaussian integration about $\rho^a$ 
through (\ref{raw}) we obtain an effective action with an auxiliary scalar
field $\rho$
\begin{equation}
S_{cs} = \int d^{p+1}\sigma \left[ \frac{1}{4\rho} ( \g_{00} - G_{0a}
\hat{G}^{ab}G_{b0} ) - \rho T^2 \det\hat{G}_{ab} + L_{WZ} \right].
\label{efs}\end{equation}
This effective action seems quadratic in the time derivative of the 
canonical variables but it should be noted that there are couplings between
the auxiliary scalar field and the canonical variables. Indeed the 
non-Gaussian integration over the remaining Lagrange multiplier field $\rho$
generates the non-linear square root structure and reproduces the starting
super D-p-brane action (\ref{act}).

In Ref.\cite{APS} for the super IIA D-p-brane the $\kappa$-symmetry is used 
to eliminate half of the components of the $\ta$ coordinates. The gauge 
choice $\ta_2 = 0, \ta_1 = \la$ makes the thory surprisingly simple, 
where the WZ term vanishes. Here we start with the 
$\kappa$-symmetry-gauge-fixed action
\begin{eqnarray}
S &=&  -T \int d^{p+1}\sigma \sqrt{-\det(\g_{ij} + {\F}_{ij})}, \nonumber \\
\g_{ij} &=& \eta_{\mu\nu} \pi^{\mu}_i \pi^{\nu}_j, \hspace{1cm} \pi^{\mu}_i =
\p_iX^{\mu} - \bar{\la}\G^{\mu}\p_i{\la}, \nonumber \\
{\F}_{ij} &=& F_{ij} + \bar{\la}\G_{\mu}\p_i\la \p_jX^{\mu} - \bar{\la}
\G_{\mu}\p_j\la\p_iX^{\mu}, 
\label{sk}\end{eqnarray}
which is invariant under the reparametrization symmetry. The extended 
Hamiltonian is expressed as
\begin{eqnarray}
H &=& \pi^a \p_a A_0 + \xi \pi^0 +   \rho T_0 
+ \rho^a T_a   - \bar{\Phi}_{\la}\psi_{\la}, \nonumber \\
\bar{\Phi}_{\la} &=& \bar{P}_{\la} + \bar{\la}[ \tilde{P}_{\mu} - 
\pi^a ( \pi_{a\mu} + \bar{\la}\G_{\mu}\p_a\la ) ]\G^{\mu} 
\label{ham}\end{eqnarray}
in terms of the corresponding reduced constraints,
where $\tilde{\pi}^a=\pi^a, \tilde{P}_{\mu} = P_{\mu} + \pi^a\bar{\la}
\G_{\mu}\p_a\la$ and $\bar{\Phi}_{\la}$ is the second class constraint. We
will analyze the equations of motion for the canonical variables.
The $\tau$ evolutions of the bosonic coordinates $X^{\mu}$ and the conjugate
momenta $P_{\mu}$ are solely generated by the constraints as 
\begin{eqnarray}
\dot{X}^{\mu} &=& \rho^a \pi^{\mu}_a + 2\rho \tilde{P}^{\mu} - \bar{\la}
\G^{\mu}\psi_{\la},  \label{xdot}  \\
\dot{P}_{\mu} &=& \p_a [ \rho^a\tilde{P}_{\mu} + ( \rho^b\pi^a - 
\rho^a\pi^b )\bar{\la}\G_{\mu}\p_b\la + 2\rho\pi^a \pi_{b\mu} \pi^b
\nonumber \\ 
&+& \rho T^2\det \hat{G}_{ab}( \hat{G}^{(a,b)} \pi_{b\mu} + 
\hat{G}^{[a,b]} \bar{\la}\G_{\mu}\p_b \la ) + \pi^a \bar{\la} \G_{\mu}
\psi_{\la} ], 
\label{pdot}
\end{eqnarray}
where $\psi_{\la}$ is set to $-\dot{\la}$ through the equations of motion
for $\la$ that will be analyzed below. The equation (\ref{pdot}) into which
the $\tilde{P}^{\mu} = ( \pi^{\mu}_0 - \rho^a\pi^{\mu}_a )/2\rho$
obtained from (\ref{xdot}) is inserted, can coincide with the Euler-Lagrange
equation of (\ref{sk}) only when 
\begin{equation}
\frac{1}{2\rho} = \frac{T\det \hat{G}_{ab}}{\sqrt{-\det G}}, \hspace{1cm}
\rho^a = - \frac{\Delta_{0a} + \Delta_{a0}}{2\det \hat{G}_{ab}}.
\label{mul}\end{equation}
Here $\Delta_{ij}$ is cofactor of the matrix $G_{ij}$ as defined by
$G^{ij} = \Delta_{ji}/\det G$. This agreement is shown by using the 
symmetric and anti-symmetric relations
\begin{eqnarray}
\hat{\Delta}_{(a,b)}\det G &=& \Delta_{(a,b)}\det \hat{G}_{ab} - ( 
\Delta_{0a}\Delta_{b0} + \Delta_{0b}\Delta_{a0} ),
\label{dels}\\
\hat{\Delta}_{[a,b]}\det G &=& \Delta_{[a,b]}\det \hat{G}_{ab} + ( 
\Delta_{0a}\Delta_{b0} - \Delta_{0b}\Delta_{a0} ),
\label{dela}\end{eqnarray}
where $\hat{\Delta}_{ab}$ is similarly defined by $\hat{G}^{ab} = 
\hat{\Delta}_{ba}/\det\hat{G}_{ab}$. The (\ref{dels}) and (\ref{dela})
relations are associated with $\hat{G}^{(a,b)}$ and $\hat{G}^{[a,b]}$ in
(\ref{pdot}) respectively. We can verify a single formula
\begin{equation}
\hat{\Delta}_{ab}\det G = \Delta_{ab}\det\hat{G}_{ab} - \Delta_{a0}
\Delta_{0b}, 
\label{delt}\end{equation}
which leads to the above two relations. In Ref. \cite{KH} for the canonical
formulation of super IIB D-p-brane the same choice as (\ref{mul}) has been
presented. The canonical equations of motion for the BI gauge field are 
given by
\begin{eqnarray}
\dot{A}_a &=& \rho^b{\F}_{ba} + 2\rho\g_{ab}\pi^b + \p_a A_0 + 
\bar{\la}\G_{\mu}\psi_{\la}\p_a X^{\mu} \nonumber \\
& & + ( \rho^b \pi^{\mu}_b + 2\rho\tilde{P}^{\mu} - \bar{\la}\G^{\mu}
\psi_{\la} ) \bar{\la}\G_{\mu}\p_a \la,  \label{adot} \\
\dot{\pi}^a &=& \p_b( \rho^b \pi^a - \rho^a \pi^b + \rho T^2\det\hat{G}_{ab}
\hat{G}^{[a,b]} ). \label{pidot}
\end{eqnarray}
In (\ref{pidot}) the elimination of momentum $\pi^a = \hat{\g}^{ab}(
{\F}_{0b} - \rho^c{\F}_{cb})/2\rho$ which is provided by combining 
(\ref{adot}) and (\ref{xdot}), 
yields an equation of motion in the configuration space which
agrees with the Euler-Lagrange equation of (\ref{sk}), where the choice
(\ref{mul}) and the anti-symmetric relation (\ref{dela}) are used.
Furthermore the canonical equations of motion for the 16-component spinor
$\la$ are written down as
\begin{eqnarray}
\dot{\la} &=& -\psi_{\la}, \nonumber \\
\dot{\bar{P}}_{\la} &=& -(\tilde{P}^{\mu} - \pi^a\p_a X^{\mu})
\bar{\psi}_{\la}\G_{\mu} + \p_a\bar{\la}\G_{\mu}J^{a\mu} + 
\p_a (\bar{\la}\G_{\mu}J^{a\mu}),
\label{lam}\end{eqnarray}
where 
\begin{eqnarray}
J^{a\mu} &=&  - \rho^a\tilde{P}^{\mu}
 + (\rho^a\pi^b - \rho^b\pi^a)\p_b X^{\mu} 
- \rho T^2\det\hat{G}_{ab}(\hat{G}^{(a,b)}\pi^{\mu}_b +
\hat{G}^{[a,b]}\p_b X^{\mu}) \nonumber \\
&-& 2\rho\pi^a\pi^b\pi^{\mu}_b + (\rho^b\pi^{\mu}_b + 
2\rho\tilde{P}^{\mu} - \bar{\la}\G^{\mu}\psi_{\la})\pi^a. 
\label{ja}\end{eqnarray}
They combine with (\ref{xdot}) to produce the Euler-Lagrange equation of 
(\ref{sk}) for $\la$ through (\ref{mul}) and both (\ref{dels}) and 
(\ref{dela}).

Now instead of the static gauge we  impose a convenient gauge for fixing
the reparametrization symmetry 
\begin{eqnarray}
\g_{00} - {\F}_{0a}\hat{\g}^{ab}{\F}_{b0} + \det(\hat{\g}+\hat{\F})_{ab} = 0,
\label{gf} \\
\g_{0a} - {\F}_{0b}\hat{\g}^{bc}{\F}_{ca} = 0, 
\label{gfa}\end{eqnarray} 
that is associated with the p+1 first class constraints $T_0 = T_a = 0$.
On setting ${\F}_{ij} = 0$ in (\ref{gf}) and (\ref{gfa}) we have a covariant
gauge fixing that was taken for the membrane theory \cite{FK,FO}.
This is considered as a natural extension of the conformal gauge in the 
string theory. Under the relation (\ref{gfa}) it is not difficult to prove
that 
\begin{equation}
G^{(0,a)} = G^{0a} + G^{a0} = 0.
\end{equation}
Further there is a useful formula
\begin{equation}
\g_{00} - G_{0a}\hat{G}^{ab}G_{b0} = K_{00} - K_{0a}\hat{K}^{ab}K_{b0}
\label{gk}\end{equation}
with $K_{ij} = \g_{ij} - {\F}_{ia}\hat{\g}^{ab}{\F}_{bj}$ used previously.
Therefore the gauge fixing reexpressed by $K_{00} + \det\hat{G}_{ab} = 0,
 K_{0a} = 0$ leads to $\g_{00} - G_{0a}\hat{G}^{ab}G_{b0} + 
\det\hat{G}_{ab} = 0$. Hence from (\ref{mul}) we deduce that
\begin{equation}
\rho = \frac{1}{2T}, \hspace{1cm} \rho^a = 0.
\label{rg}\end{equation}
In this gauge the canonical equations of motion for the Hamiltonian 
(\ref{ham}) with (\ref{rg}) are considerably 
simplified as
\begin{equation}
P_{\mu} = T\pi_{0\mu} - \pi^a\bar{\la}\G_{\mu}\p_a\la, \hspace{1cm} \pi^a =
T\hat{\g}^{ab}{\F}_{0b}
\label{pp}\end{equation}
with $\tilde{P}_{\mu} = T\pi_{0\mu}$ and 
\begin{eqnarray}
\dot{P}_{\mu} &=& \p_a[ \pi^a(\frac{1}{T}\pi_{b\mu}\pi^b - \bar{\la}\G_{\mu}
\dot{\la} ) + \frac{T}{2}\det \hat{G}_{ab}(\hat{G}^{(a,b)}\pi_{b\mu} + 
\hat{G}^{[a,b]}\bar{\la}\G_{\mu}\p_b\la)], \nonumber \\
\dot{\pi}^a &=& \frac{T}{2} \p_b (\det\hat{G}_{ab} \hat{G}^{[a,b]} ), 
\label{ppi}\end{eqnarray}
which are read off directly from (\ref{xdot}), (\ref{pdot}), (\ref{adot}) 
and (\ref{pidot}). 
Choosing $\rho$ and $\rho^a$ as (\ref{rg}) in the extended Hamiltonian is
equivalent to choosing the conformal-like gauge (\ref{gf}), (\ref{gfa}).
Then through the simple forms of momenta (\ref{pp}), 
 the first class constraints $T_0$ and $T_a$ turn out to be
the gauge-fixing expressions of (\ref{gf}) and (\ref{gfa}) respectively.
The canonical equations of motion for $\la$ in the covariant gauge are also
obtained from (\ref{lam}) and (\ref{ja}). The equations in 
(\ref{ppi}), which are rewritten in the configuration space by substitution
of (\ref{pp}), again agree with the Euler-Lagrange
equations for the super D-p-brane action (\ref{sk}) under the 
conformal-like gauge fixing
\begin{eqnarray}
\frac{\p}{\p \tau}( T\pi_{0\mu} - \pi^a\bar{\la}\G_{\mu}\p_a\la )
- \frac{T}{2}\p_a [ \sqrt{-\det G}( G^{(a,i)}\pi_{i\mu} + G^{[a,i]}\la
\G_{\mu}\p_i\la ) ] = 0, \nonumber \\
\frac{\p}{\p \tau}\pi^a - \frac{T}{2}\p_b ( \sqrt{-\det G}G^{[a,b]} ) = 0.
\end{eqnarray}
Gathering together we find an effective simplified action
\begin{equation}
S = \frac{T}{2}\int d^{p+1}\sigma ( \pi^{\mu}_0 \pi_{0\mu} + {\F}_{0a}
\hat{\g}^{ab}{\F}_{0b} - \det(\hat{\g} + \hat{\F})_{ab})
\label{ss}\end{equation}
with $\pi^{\mu}_0 = \dot{X}^{\mu} - \bar{\la}\G^{\mu}\dot{\la}$, 
whose Euler-Lagrange equations reproduce (\ref{ppi}) and the equation
of motion for $\la$. It is interesting to note that the resulting action
(\ref{ss}) is directly obtained from the configuration effective action
(\ref{efs}) if we choose $\rho = 1/2T, \ta_1 = \la, \ta_2 = 0$ and use
(\ref{gk}) with $K_{0a} = 0$. Compared with the non-linearity of the 
starting super D-p-brane action in the square root form 
the obtained action is mainly quadratic in the time derivative of the
bosonic ten-dimensional coordinates and the BI gauge field.
The third term in (\ref{ss}) is generally non-quadratic but in the
super IIA D-2-brane case it is decomposed into a quadratic part for the
BI gauge field and a standard non-quadratic determinant part for the 
bosonic coordinates. Moreover, if we begin with the linearized action 
(\ref{ss}), since the velocities such as $\dot{X}^{\mu}$ and 
$\dot{A}_a$ now can be described in terms of the canonical variables,
we have a non-zero canonical Hamiltonian $( \tilde{P}^2_{\mu} + \pi^a\g_{ab}
\pi^b + \det(\hat{\g}+\hat{\F})_{ab})/2T + \p_a A_0\pi^a$ 
with the fermionic and
BI gauge symmetry constraints and then go back to the full Hamiltonian
(\ref{ham}) accompanied with (\ref{rg}). This Hamiltonian that is 
quadratic in the momenta of bosonic variables has the similar 
structure to the light-cone Hamiltonians for the M-2-brane \cite{WHN} and 
also the D-p-brane \cite{JL,MMM}.

In conclusion for the $\kappa$-symmetric IIA D-p-brane 
from the phase space action  
 which is constructed by using the Dirac Hamiltonian we have
recovered the non-linear BING action with the WZ term through equations of
motion for the canonical momenta and the Lagrange multiplier fields. 
For this classical equivalence where the dynamics is generated by the 
constraints it is important to treat the fermionic constraints 
properly. On the way of this demonstration we have extracted an 
effective action which is specified by an auxiliary scalar field. 
At first sight it seems quadratic in the time derivative of the 
canonical variables. But it cannot be said that we obtain a truely
linearized effective action, because the non-linearity is concealed in
the coupling of the auxiliary scalar field to the canonical variables.
For the $\kappa$-symmetry-gauge-fixed action we have found a suitable
conformal-like covariant gauge that certainly linearizes the square root
structure of BING action and generates a simplified effective action.
Comparing the two effective actions we have seen that the conformal-like
gauge fixing corresponds to the setting of the auxiliary scalar field at
a particular value. This setting instead of integrating surely linearizes
the BING action. To the same level as the light-cone gauge prescription
the resulting covariant action has been simplified. 
The linearized simple form of our effective action allows us to hope 
that it opens the way to study the covariant quantization as well as
spectrum of super IIA D-branes.

\end{document}